\documentclass[12pt,a4paper]{article}
\usepackage[utf8]{inputenc}
\usepackage[plainpages=false,pdfpagelabels=true]{hyperref}
\usepackage{amssymb,amsthm}
\usepackage[margin=1.5 in]{geometry}
\usepackage{graphicx}
\usepackage{a4wide}  
\usepackage{amsfonts}
\usepackage{amsmath}
\usepackage{bbm}
\usepackage{booktabs} 
\usepackage{multicol}
\usepackage{ifpdf}
\ifpdf

\hypersetup{%
  pdftitle    = {Dynamics in Wormhole spacetimes  and the Jacobi metric approach}
  pdfkeywords = {Wormholes, Jacobi metric, geodesics},
  pdfauthor   = {Oscar},
  plainpages  = true,
  colorlinks  = true,
  citecolor   = blue,
  urlcolor    = blue,
  linkcolor   = black
}

\newcommand{\arxiv}[1]{{\tt
\href{http://www.arXiv.org/abs/#1}{#1}}}
\newcommand{\doi}[1]{{\tt DOI:\href{http://dx.doi.org/#1}{#1}}}

\makeatletter
\@addtoreset{equation}{section}
\makeatother

\begin{document}

\begin{center}

{\Large {\bf Dynamics in wormhole spacetimes: a  Jacobi metric approach}}

\vspace{1.5cm}

\renewcommand{\thefootnote}{\alph{footnote}}
{\sl\large  Marcos Argañaraz\footnote{E-mail: marcos.arganaraz [at] unc.edu.ar} and Oscar Lasso Andino}\footnote{E-mail: {oscar.lasso [at] udla.edu.ec}}

\setcounter{footnote}{0}
\renewcommand{\thefootnote}{\arabic{footnote}}

\vspace{1.5cm}

{\it $^a$Facultad de Matemática, Astronomía, Física y Computación, Universidad Nacional de Córdoba, Instituto de Física Enrique Gaviola, CONICET, (5000) Córdoba, Argentina.}
\ \vspace{0.3cm}

{\it $^b$ Escuela de Ciencias Físicas y Matemáticas, Universidad de Las Américas,\\
C/. José Queri, C.P. 170504, Quito, Ecuador}\\ \vspace{0.3cm}

\vspace{1.8cm}


{\bf Abstract}

\end{center}

\begin{quotation}
This article deals with the study of the dynamics of particles in different wormhole geometries. Using the Jacobi metric approach we study the geodesic motion on the Morris-Thorne wormhole. We found the only stable circular orbit located at the throat. We show that the Gaussian curvature of the Jacobi metric is directly related with the wormhole flare-out condition. We provide a simple test for determining the existence of a throat in a spacetime by using the Gaussian curvature of the associated Jacobi metric only. We discuss about the trajectories in the Kepler problem in a wormhole background. Finally, we discuss about the restrictions over the stress-energy tensor imposed by the existence of elliptic orbits in the Kepler problem. 
\end{quotation}

\newpage
\pagestyle{plain}


\newpage


\section{Introduction}

In 1935 Einstein and Rosen introduced the idea of wormholes. However, it was not until Morris and Thorne published their article  \cite{Morris:1988cz} that the study of wormholes explode. These wormholes where supposed to be traversable but only if we accept the existence of certain type of matter, what they called ``exotic" matter. This kind of matter  violates the  energy conditions and therefore it is considered unphysical\cite{Martin-Moruno:2013wfa, Lobo:2017oab}. There has been a lot of effort in building wormholes sustained by normal matter. For instance, in \cite{Kuhfittig:2018voi} it is proposed that the existence of an extra dimension is the responsible for the violation of the null energy condition. In \cite{Hammad:2018ydd} it is shown that the  wormhole can be traversable in spacetimes with torsion. See also  \cite{Garattini:2007fe, Barros:2018lca,Bhar:2016vdn}. There are some successful attempts of traversability in thin-shell wormholes for alternative theories of gravity \cite{Eiroa:2008hv,Richarte:2007zz,Richarte:2010bd,Maeda:2008nz}. Wormholes with different asymptotic limits have also been studied. Recently, a novel traversable wormhole solution has been found. This solution is asymptotically locally AdS and it is a solution to Einstein equations with a cosmological constant\cite{Anabalon:2018rzq}. 
There are  also studies of wormholes in the context of the Noncommutative geometry, which seems to be a promising branch in the wormhole community \cite{Rahaman:2014dpa, Abreu:2012fg,Kuhfittig:2013ib,Garattini:2008xz}. In \cite{Myrzakulov:2015kda} they found a traversable wormhole in an alternative theory of gravity called mimetic gravity, closely related to noncommutative geometry theories. This kind of wormhole, in Genereal Relativity become an Ellis-Bronnikov wormhole type. \\ 
More recently, it has been discovered that certain kind of long traversable wormholes are allowed \cite{Maldacena:2018gjk}. These long wormholes, that connect two different regions of spacetime, allow to go through them, but it is faster to go from one side to the other  through the ambient space\footnote{These objects are not allowed in classical physics. The authors are able to build  a wormhole taking into account quantum effects.}.\\
The existence of wormholes is under debate. In cite \cite{Rahaman:2013xoa} the possible existence of wormholes in the galactic halo region is discussed, while in \cite{Bueno:2017hyj} the wave forms of the echoes of static and stationary, traversable wormholes are studied.  They studied the quasinormal spectrum of perturbations  in  some wormhole  spacetimes, and how the echoes coming from the signal, connected to  such spectrum, can be reconstructed from a primary signal. 

An important aspect related to traversability is the geodesic motion of point particles. What we are usually looking for are geodesics capable of tunneling through the wormhole throat. In \cite{Olmo:2015bya} the authors study the geodesic configuration of three different types of wormhole solutions: the Minkowski, the Schwarzschild, and the Reissner–Nordstrom-like wormholes. The most important conclusion is that these wormhole spacetimes allow geodesically  complete  paths, and therefore they can be traversed. Geodesic motion in Schwarzschild and Kerr thin-shell  wormholes has been studied in \cite{Kagramanova:2013mwv}.

There are different approaches for studying  wormhole geodesics.  In \cite{Muller:2008zza} the author studied the null and timelike geodesics using the typical way, namely solving the geodesic equation and writing the results in terms of the Jacobian elliptic and elliptic integral functions. The analytical results make it possible to find a geodesic connecting any two distant events in the wormhole spacetime. In \cite{Mishra:2017yrh} an  study of null and timelike geodesics in the background of wormhole geometries is presented. Dealing with static and dynamic spherically symmetric wormholes they find null geodesics and photon spheres. 

We want to pursue another direction. We look for a method that help us to study the dynamics only by knowing geometric properties. If we want to study wormholes in the context of a purely geometric theory, such as string theory, we need to follow an approach that allows us to characterize the dynamics by using geometric properties only. We want to study the dynamics of wormholes using the Jacobi metric approach \cite{Gibbons:2015qja,Chanda:2016aph,Tsiganov:2001}.\\ This approach will let us to study, for example, the Kepler problem, in an easier way (sometimes the only one). 
On the Jacobi approach there is a fixed energy restriction. The Lagrangian from which  the equations of motion are obtained  has to be restricted. This Lagrangian is going to be dependent on the Jacobi metric which involves the particle energy in an explicit way. The energy becomes an additional parameter, such as the rest mass of the particle, although  the  motion is still geodesic in the Jacobi metric space. Associated with the Jacobi metric is the Gaussian curvature, whose sign turns out to be determined by the flare-out condition. Knowing the sign of this curvature the  trajectories in the Kepler problem can be classified in the following way:  positive curvature corresponds to elliptic orbits,  negative curvature corresponds to the hyperbolic ones and zero curvature corresponds to parabolic orbits. The zones where the flare-out condition is satisfied will have a definite sign, and therefore we can identify them only by determining the sign of the Gaussian curvature.
Wormhole geometries have not been studied using this approach, and we hope that geometric properties such as the flare-out condition can be reinterpreted in a more profound or an alternative way. Indeed, we were able to use the Gauss curvature of the Jacobi metric for determining the existence of a throat in a given spacetime, leading us to define a method for determining if a spacetime is a wormhole. 

In section \ref{Marpetuistheory} we present a brief review of the Jacobi metric formalism. We describe the Jacobi metric and its properties. In section \ref{staticwormholes} we apply the Jacobi metric approach to static wormhole geometries. We find the stable circular orbits and the bound states for the wormhole. In section \ref{flareout} we show that the sign of the Gaussian curvature is defined by the flare-out condition. We also discuss about the Kepler problem and the kind of matter that allow different types of trajectories. In section \ref{discussion} we present the discussion and final comments. Finally, in appendix \ref{appendix} we present the calculated Jacobi metric for different wormhole spacetimes and a different redshift function.


\section{The Jacobi metric and the Marpetuis principle}\label{Marpetuistheory}
Here we provide an introduction to the Jacobi metric approach. We will use this approach to study the dynamics on different types of wormholes.\\
Given a $n-$dimensional Lagrangian $\mathcal{L}(q_{i},\dot{q}_{i})$ ,where $q_{i}$ is a generalized coordinate, the extremals  of its correspondent action $S=\int{\mathcal{L}(q_{i},\dot{q}_{i})}$ are also the extremals of a reduced action in a $(n-1)-$ dimensional phase space. This phase space is a level set of the Hamiltonian $H(q_{i},p_{i})=E$, where $p_{i}$ is the conjugated momenta of $q_{i}$ .\\
In general, let us define a Hamiltonian in a Riemannian manifold with metric $g_{ij}$
\begin{equation}
H(q,p)=g^{ij}(q)p_{i}p_{j}+V(q).
\end{equation}
In a fixed energy submanifold the trajectories of $H$ will be the same trajectories as the ones for an alternative Hamiltonian given by
\begin{equation}\label{hamiltonian:1}
\bar{\mathcal{H}}=\frac{g^{ij}(q)}{E-V(q)}p_{i}p_{j},
\end{equation}
together with the transformation
\begin{equation}\label{parameter:1}
d\bar{t}=(E-V(q))dt.
\end{equation}
The transformation given in (\ref{hamiltonian:1}) and (\ref{parameter:1}) is a Jacobi transformation.
This new Hamiltonian has  properties that are attractive. For instance, it encodes geodesic motion information, allowing to study dynamical problems. It also let us to classify all trajectories using the Gauss curvature induced by the Jacobi metric.\\

Let us consider a Lagrangian of the form
\begin{equation}\label{lagrangian}
L=\frac{1}{2}m_{ij}\dot{x}^{i}\dot{x}^{j}-V(x),
\end{equation}
where $m_{ij}$ is a mass matrix depending on space coordinates.
The constrained motion ($H(q_{i},p_{i})=E$) of a particle is described by the geodesics of a re-scaled metric
\begin{equation}\label{jacobi}
J_{ij}dx^{i}dx^{j}=2(E-V)m_{ij}dx^{i}dx^{j}.
\end{equation}
According to the Marpetuis principle, the trajectories of the system described by the constancy of (\ref{lagrangian}) are the geodesics of the metric (\ref{jacobi}). This fact  has profound geometrical implications. The metric (\ref{jacobi}) is called the Jacobi metric. \\
The Jacobi metric formalism has been used by Gibbons et al. \cite{Gibbons:2015qja} to study the dynamics of a massive particle in a Schwarzschild spacetime.  The free motion of a particle in a static spacetime can be described by a energy dependent metric, which turns out to be a Riemannian metric on the spatial sections, in clear analogy to the classical case. The Kepler problem has been studied in \cite{Chanda:2016sjg}. Recently, the motion of massive particles in the Reissner-Nordström spacetime have been studied in \cite{Das:2016opi}. For a good introduction see \cite{Szydlowski:1996}
Thus, for Lorentzian geometries, the temporal part is going to behave as the potential in the equation (\ref{lagrangian}). We start with static spacetimes. We consider a Lorentzian metric written as
\begin{equation}
ds^{2}=-V^{2}dt^{2}+g_{ij}dx^{i}dx^{j},
\end{equation}
where $V=V(r)$ is a continuous function of $r$. The action for a massive particle becomes
\begin{equation}
S=-m\int dt\sqrt{V^{2}-g_{ij}\dot{x}^{i}\dot{x}^{j}},
\end{equation}
where, as usual, $\dot{x}^{i}=\frac{dx^{i}}{dt}$. The canonical momentum is 
\begin{equation}
p_{i}=\frac{mg_{ij}\dot{x}^{j}}{\sqrt{V^{2}-g_{ij}\dot{x}^{i}\dot{x}^{j}}},
\end{equation}
which leads to the hamiltonian
\begin{equation}
H=\sqrt{V^{2}g^{ij}p_{i}p_{j}+m^{2}V^{2}}.
\end{equation}
Using $p_{i}=\partial_{i}S$ we can write the Hamilton-Jacobi equation as
\begin{equation}\label{HJ:1}
\sqrt{V^{2}g^{ij}\partial^{i}S\partial_{j}S+m^{2}V^{2}}=E.
\end{equation}
Defining $f_{ij}=V^{-2}g_{ij}$ (the Fermat metric) the eq.(\ref{HJ:1}) can be expressed as
\begin{equation}\label{HamiltonJeq}
\frac{1}{E^{2}-m^{2}V^{2}}f^{ij}\partial_{i}S\partial_{j}S=1,
\end{equation}

Equation (\ref{HamiltonJeq}) is the Hamilton-Jacobi equation for the geodesics of the Jacobi metric $J_{ij}$ given by
\begin{equation}\label{jacobi metric}
J_{ij}dx^{i}dx^{j}=(E^{2}-m^{2}V^{2})V^{-2}g_{ij}dx^{i}dx^{j}.
\end{equation}

For massless particles ($m=0$), the Jacobi metric becomes the Fermat metric up to a factor of $E^{2}$, and therefore the geodesics do not depend upon $E$. In the massive case ($m\neq 0$) the dependence of the geodesics upon the energy $E$ will appear \cite{Gibbons:2015qja}.

Throughout the article we will compare, whenever possible, the results from the Jacobi metric approach with the results obtained using the conventional approach.\\ 
There is a similar result in the second-order variational calculus. In \cite{Izquierdo:2002jt} it is shown that the Morse theory associated to the dynamical problem is the Morse theory associated to the Jacobi metric. This is a result that have not been exploited, specially in the field of wormholes.

\section{Static wormhole geometries and the Jacobi metric}\label{staticwormholes}
There are different ans\"{a}tze for metrics describing wormhole geometries. The simplest one is the Morris-Thorne wormhole:
\begin{equation}\label{wormhole}
ds^{2}=-e^{2\Phi(r)}dt^{2}+\frac{dr^{2}}{1-\frac{b(r)}{r}}+r^{2}(d\theta^{2}+\sin^{2}(\theta) d\phi^{2}),
\end{equation} 
where $\Phi(r)$ is the redshift function, and $b(r)$ is the shape function. It describes a horizonless  asymptotically flat spherically symmetric spacetime  \cite{Morris:1988cz}. The metric (\ref{wormhole}) is a solution of the Einstein equations provided that the stress-energy tensor satisfies certain constraints.\\
The radial coordinate $r$ has a minimum value, it decreases from $+\infty$ to $b(r_{o})=r_{o}$, and then, it goes from  $r_{o}$ to $-\infty$. The throat of the wormhole is located at $r_{o}$. The $g_{rr}$ component of the metric diverges at the throat. Moreover, the proper distance $l(r)$ is required to be finite everywhere.\footnote{The proper distance is defined by $l(r)\pm \int_{r_{o}}^{r}\left(1-\frac{b(r)}{r}\right)$. It decreases form $+\infty$ to $0$ (in the throat),  and then, it decreases form $0$ to $-\infty$} In order to ensure the traversability of the wormhole it is required that $g_{tt}\neq 0$, implying that $\Phi(r)$ must be finite \cite{Lobo:2017oab}.\\
Using (\ref{jacobi metric}) it is straightforward to find the Jacobi metric corresponding to (\ref{wormhole}). It is given by
\begin{equation}\label{jacobim:1}
J_{ij}dx^{i}dx^{j}=\left(E^{2}-m^{2}e^{2\Phi(r)}\right)\left(\frac{dr^{2}}{\left(1-\frac{b(r)}{r}\right)e^{2\Phi(r)}}+\frac{r^{2}d\theta^{2}+r^{2}\sin^{2}(\theta)d\phi^{2}}{e^{2\Phi(r)}}\right).
\end{equation} 
The Jacobi metric (\ref{jacobim:1}) is defined in the spatial sections of the original metric\footnote{ 
The quantity in the first bracket of (\ref{jacobim:1}) is the conformal factor and the Fermat metric is 
\begin{equation}\label{fermat}
F_{ij}=\left(\frac{dr^{2}}{\left(1-\frac{b(r)}{r}\right)e^{2\Phi(r)}}+\frac{r^{2}d\theta^{2}+r^{2}\sin^{2}(\theta)d\phi^{2}}{e^{2\Phi(r)}}\right).
\end{equation}} (\ref{wormhole})

Without loss of generality -because of spherical symmetry- we can consider the sections of the metric when $\theta=\frac{\pi}{2}$, then the equation (\ref{jacobim:1}) becomes
\begin{equation}\label{restrictedm:1}
ds^{2}=\left(E^{2}-m^{2}e^{2\Phi(r)}\right)\left(\frac{dr^{2}}{\left(1-\frac{b(r)}{r}\right)e^{2\Phi(r)}}+\frac{r^{2}d\phi^{2}}{e^{2\Phi(r)}}\right).
\end{equation}
The Clairaut constant corresponds to the angular momentum, which is constant, thus
\begin{equation}\label{clairaut:1}
l=\left(E^{2}-m^{2}e^{2\Phi(r)}\right)r^{2}e^{-2\Phi(r)}\left(\frac{d\phi}{ds}\right),
\end{equation}
then
\begin{equation}
\left(E^{2}-m^{2}e^{2\Phi(r)}\right)\left(\frac{1}{\left(1-\frac{b(r)}{r}\right)e^{2\Phi(r)}}\left(\frac{dr}{ds}\right)^{2}+\frac{r^{2}}{e^{2\Phi(r)}}\left(\frac{d\phi}{ds}\right)^{2}\right)=1.
\end{equation}
Using (\ref{restrictedm:1}) and  (\ref{clairaut:1}) we find
\begin{equation}
\frac{(E^{2}-m^{2}e^{2\Phi(r)})^{2}}{e^{4\Phi}}\left(\frac{dr}{ds}\right)^{2}=\left(\frac{E^{2}}{e^{2\Phi}}-m^{2}-\frac{l^{2}}{r^{2}}\right)\left(1-\frac{b(r)}{r}\right),
\end{equation}
Making the change of variable 
\begin{equation}
d\tau=\frac{m e^{2\Phi}}{E^{2}-m^{2}e^{2\Phi}}ds
\end{equation}
we get the known result, (see eq.(36) in \cite{Mishra:2017yrh}):
\begin{equation}\label{radialeq:1}
m^{2}\left(\frac{dr}{d\tau}\right)^{2}=\left(\frac{E^{2}}{e^{2\Phi}}-m^{2}-\frac{l^{2}}{r^{2}}\right)\left(1-\frac{b(r)}{r}\right)
\end{equation}
It is common to set $u=\frac{1}{r}$, and therefore, $u$ will satisfy the Binet's equation
\begin{equation}\label{firstordeq:1}
\left(\frac{du}{d\phi}\right)^{2}=\frac{1}{h^{2}}\left(\frac{\tilde{E}^{2}}{e^{2\Phi(1/u)}}-1-h^{2}u^{2}\right)(1-b(1/u)u),
\end{equation}
where we have set $\tilde{E}=\frac{E}{m}$ and $h=\frac{l}{m}$. We have found a first integral  depending on the $\Phi$ and $b$ functions. We take the shape function \footnote{The case $n=2$ is the Morris-Thorne wormhole}
\begin{equation}\label{shapef}
b(r)=b_{o}^{n}r^{1-n}\,\,\, ,n>0.
\end{equation}
which in terms of the variable $u$ is
\begin{equation}\label{shapef2}
b(u)=b_{o}^{n}u^{n-1},
\end{equation}

then, the equation (\ref{firstordeq:1}) transforms to
\begin{equation}
\left(\frac{du}{d\phi}\right)^{2}=\frac{1}{h^{2}}\left(\frac{\tilde{E}^{2}}{e^{2\Phi(\frac{1}{u})}}-1-h^{2}u^{2}\right)(1-b_{o}^{n}u^{n}).
\end{equation}
After some massage we arrive to
\begin{equation}\label{firstorder}
\left(\frac{du}{d\phi}\right)^{2}=b_{o}^{n}\left(u^{n+2}-\frac{1}{h^{2}}\left(\frac{\tilde{E}^{2}}{e^{2\Phi}}-1\right)u^{n}-\frac{u^{2}}{b_{o}^{n}}+ \frac{1}{h^{2}b_{o}^{n}}\left(\frac{\tilde{E}}{e^{2\Phi}}-1\right)\right).
\end{equation}

The previous equation is a first integral of motion, and it is one of our important results. However, if we want to solve the equation we have to provide a redshift function $\Phi$. As a first example we take $\Phi=0$. See appendix \ref{anotherf} for the calculation using another redshift function.

\subsection{The simplest redshift function: $\Phi=0$}\label{stableorbit}

Here we study the dynamics of the wormhole with the redshift function given by  $e^{\Phi}=1$. Hence, the equation (\ref{firstorder}) becomes
\begin{equation}\label{psieq:1}
\left(\frac{du}{d\phi}\right)^{2}=b_{o}^{n}\left(u^{n+2}-C u^{n}-\frac{u^{2}}{b_{o}^{n}}+ \frac{C}{b_{o}^{n}}\right),
\end{equation}

where we have set $C=\frac{\tilde{E}^{2}-1}{h^{2}}$ and $h=\frac{l}{m}$. Now, we study the circular orbits for a particle of mass $m$.

\subsubsection{Circular orbits}
If we define $f(u)=b_{o}^{n}\left(u^{n+2}-C u^{n}-\frac{u^{2}}{b_{o}^{n}}+ \frac{C}{b_{o}^{n}}\right)$ we can study the existence of circular orbits. In order to have circular orbits the function $f(u)$ needs to satisfy  two requisites, namely $f(u)$ and $f'(u)$ should become zero at some ($u=u_{c}$). Hence, we set
\begin{eqnarray}
f(u)&=&0, \label{fu:1}\\
f'(u)& = &b_{o}^{n}\left((n+2)u^{n+1}-nCu^{n-1}-\frac{2u}{b_{o}^{n}}\right)=0.\label{fu:2}
\end{eqnarray}

Using (\ref{fu:1}) and (\ref{fu:2}) we find that for a  circular orbit, the energy per unit mass $\tilde{E}$ and the momentum per unit mass $h$ have to satisfy 
\begin{equation}\label{cfinal}
\frac{\tilde{E}^2-1}{h^2}=\frac{1}{b_{o}^2}
\end{equation}

Contrary to what happens in the Schwarzchild and Reissner- Nördstrom black hole cases we have that $\frac{\tilde{E}^2-1}{h^2}$ is independent of $u$, and therefore we will have only one circular orbit, which is going to be located at the throat. Indeed, from equation (\ref{fu:2}) we find
\begin{equation}\label{cfunction}
C=\frac{u_{c}^2}{n}\left(2+n-\frac{2}{b_{o}^n u_{c}^n}\right).
\end{equation}

After replacing (\ref{cfunction}) in (\ref{fu:1}) we obtain 
\begin{equation}\label{circular}
b_{o}^n u_{c}^n-1=0,
\end{equation}
which clearly implies that $1/u_{c}=r_{c}=b_{o}$. 
The previous result says that for a given $\tilde{E}$ and  $h$, there is only one circular orbit, and it is located at the throat.
Moreover, we want to know if this circular orbit is stable. The stability criteria is $f''(u)>0$. Hence, a stable orbit will be reached when:
\begin{equation}\label{fu:3}
f''(u) = b_{o}^n\left((n+2)(n+1)u_{c}^{n}-C(n-1)nu_{c}^{n-2}-\frac{2}{b_{o}^{n}}\right)>0.
\end{equation}
In the wormhole geometries (\ref{wormhole}), 
with redshift function $\Phi=0$ and shape function  given by (\ref{shapef2}) the only circular orbit will satisfy
\begin{equation}
f''(u=u_{c},C=\frac{1}{b_{o}^2})=4n>0.
\end{equation}
From the previous equation we conclude that the unique circular orbit located at the throat is stable\footnote{In our wormhole, with the only stable orbit at the throat, it does not make any sense to look for the (ISCO), the Innermost Stable Circular Orbit, the boundary orbit on which the finite motion is still possible. The condition for the (ISCO) is $f''(u)=0$.}. Later on, we will see that this conclusion has strong implications and it is related with the fact that apart form the throat orbit all the orbits are open.\\

\textbf{$\bullet$ Bound states for The Morris-Thorne wormhole (n=2)}\\

The equation (\ref{psieq:1}) can be solved for any $n$. Nevertheless, it is going to be really difficult to solve it for higher  $n$ ($n>3$). Again, we take $n=2$, the Morris-Thorne wormholes, then equation (\ref{psieq:1}) becomes

\begin{equation}\label{firstordeq:2}
\left(\frac{du}{d\phi}\right)^{2}=b_{o}^{2}\left(u^{4}-\left(C+\frac{1}{b_{o}^{2}}\right)u^{2}+\frac{C}{b_{o}^{2}}\right).
\end{equation}

Let us try to write the solution to (\ref{sol}) in terms of the roots of the equation.
The integral of (\ref{firstordeq:2}) is
\begin{equation}\label{firstordeq:3}
\left(\frac{du}{d\phi}\right)^{2}=b_{o}^{2}\left(u^{4}-\left(C+\frac{1}{b_{o}^{2}}\right)u^{2}+\frac{C}{b_{o}^{2}}\right)=b_{o}^{2}(u-\alpha)(u-\beta)(u-\gamma)(u-\delta),
\end{equation}
with
\begin{eqnarray}\label{restrictions:1}
\alpha+\beta+\gamma+\delta=0\,\,\,\,\,\,\,\,\,\,\,\,\,\,\,\,\left(\frac{1}{b_{o}^{2}}+C\right)=(\beta+\alpha)(\delta+\gamma)+\gamma\delta+\alpha\beta\\
\alpha\beta\gamma\delta=\frac{C}{b_{o}^{2}} \,\,\,\,\,\,\,\,\,\,\,\,\,\,\,\,\,\,\,\,\,\,\,\,\,\,\,\,\,\,\,\,\,\,\,\,\,\,\,\,\,\,\,\alpha\beta\delta+\alpha\beta\gamma+\beta\gamma\delta+\alpha\gamma\delta=0
\end{eqnarray}
It can be easily seen that the constants are given by
\begin{eqnarray}
\alpha= -\sqrt{C},\,\,\,\,\beta=-\frac{1}{b_{o}},\,\,\,\,\gamma=\frac{1}{b_{o}},\,\,\,\,\delta=\sqrt{C}.
\end{eqnarray}
The solution to (\ref{firstordeq:3}) can be written as a function of the Jacobi' elliptic functions\footnote{We take $\alpha >\beta\geq u \geq\gamma>\delta >0$ as one of the possible cases where $u$ can be located. If we chose a different order, because of the limits of integration, we are going to have different results.}. Thus, the equation (\ref{firstordeq:3}) can be written 
\begin{equation}
\frac{du}{\sqrt{(u-\alpha)(u-\beta)(u-\gamma)(u-\delta)}}=b_{o}^{2}d\phi.
\end{equation}
We can solve the previous equation and find an exact solution in terms of the Jacobi elliptic functions
\begin{equation}\label{sol}
r(\phi)=\frac{1}{\sqrt{C}sn(b_{o}\phi,bo^2C)},
\end{equation}
where $sn$  is the first elliptic Jacobi function, and we have assumed that $E^2>m^2$, therefore $C>0$. For the case $E^2<m^2$ we can proceed as before, however, now the equation (\ref{firstordeq:2}) will have complex roots
\begin{eqnarray}
\alpha= -i\sqrt{\lvert C\rvert},\,\,\,\,\beta=-\frac{1}{b_{o}},\,\,\,\,\gamma=\frac{1}{b_{o}},\,\,\,\,\delta= i\sqrt{\lvert C\rvert}.
\end{eqnarray}
Then, the solution to (\ref{firstordeq:2}) is given by
\begin{equation}\label{orbitsCn}
r(\phi)=b_{o}\cos\left(sn\left(\sqrt{b_{o}^2\lvert C\rvert+1}\,\,\phi,\frac{b_{o}\sqrt{\lvert C \rvert}}{\sqrt{b_{o}^2\lvert C\rvert+1}}\right) \right)
\end{equation}

\begin{figure*}
\begin{multicols}{2}
    \includegraphics[scale=0.8]{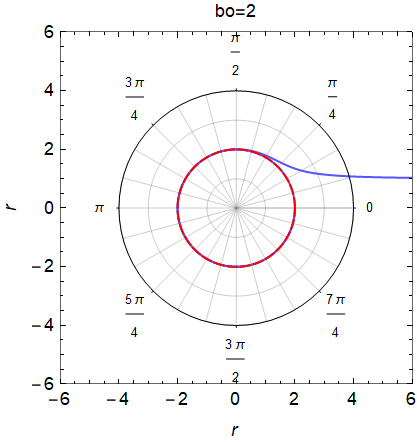}\par 
     
    \includegraphics[scale=0.8]{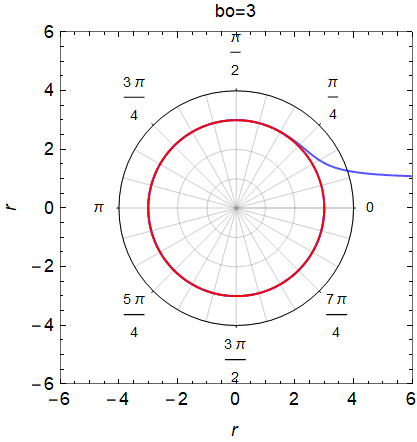}\par 
     
    \end{multicols}
    \caption{Polar plot of a trajectory of a particle coming form infinity, with different throat sizes $bo=2$ (left panel), $bo=3$ (right panel). We have used equation (\ref{sol}). The blue line represents a trajectory with the only condition $C=1/bo^2$. This graph shows the only circular stable orbit located at the throat.}
    \label{closed:1}
\end{figure*}

\begin{figure}
\begin{multicols}{2}
    \includegraphics[scale=0.8]{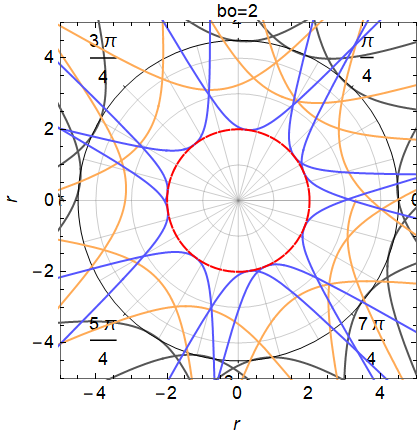}\par
     
      \includegraphics[scale=0.8]{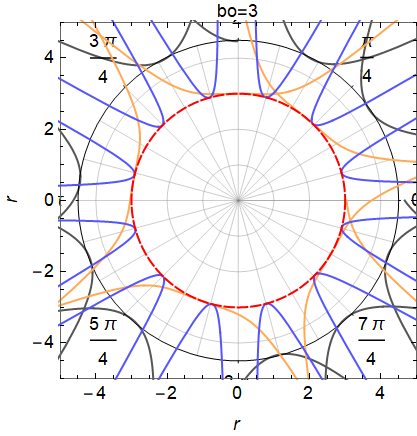}\par 
     
    \end{multicols}
    \caption{Polar plot of open trajectories of a particle coming from infinity. We have used equation (\ref{sol}) with different throat sizes $bo=2$ (left panel), $bo=3$ (right panel). The colors label different values of $C$, black is for $C=0.05$, orange is for $C=0.1$ and blue is for $C=0.5$. All trajectories are open and cannot go beyond the throat.}
    \label{open:2}
\end{figure}
In figure \ref{closed:1} the red points correspond to the throat. The size of the throat is given by $bo$. The interesting thing in both panels is that we have plotted a trajectory, in blue, that satisfies restriction \eqref{cfinal}. The blue trajectory comes from the infinity and  joins the circular stable orbit at the throat. Therefore, if a particle comes from infinity with an energy per unit mass that satisfies \eqref{cfinal}, it will be caught at the circular orbit and will remain there. We can see in both graphs the same phenomena. Only by using information coming from the Jacobi metric we are able to determine the only stable circular orbit.\\
In figure \ref{open:2} we have plotted different trajectories, and none of them satisfies (\ref{cfinal}), therefore none of them is circular neither closed. A particle coming from infinity will return to infinity. Note that the trajectories  can not go beyond the throat. These graphics support our claiming that the only stable circular  orbit is at the throat. Moreover, they provide us with a better insight about the possible type of trajectories determined by \eqref{sol}.

The equation (\ref{firstorder}) was already known. However, the Jacobi approach let us arrive to it straightforwardly. And the Jacobi metric approach method helps us to go a little bit farther. We can determine the existence of this kind of trajectories before calculating the geodesics and even before the test of extremality and minimality of the throat surface, just by calculating the Gaussian curvature of the Jacobi metric. We will see in the next section that all the points located at the throat are saddle points of the wormhole surface. In the next section we will show that the Gaussian curvature of the Jacobi metric is related with the flare-out condition of the wormhole.


\section{The flare-out condition and the Jacobi metric}\label{flareout}

We know that the Gauss curvature is crucial for studying the dynamics using the Jacobi approach. We will show that the flare-out condition is directly related with the Gaussian curvature of the Jacobi metric. This relation will lead us to establish new conditions to the matter sustaining the wormhole and the type of trajectories allowed when solving the Kepler problem.\\
The metric (\ref{restrictedm:1}) can be written as
\begin{equation}\label{metricfg}
ds^2=f(r)^{2}\left(\frac{dr^2}{g(r)^2}+r^2d\phi^2\right),
\end{equation}
where 
\begin{equation}\label{frgr}
f(r)^2=(E^2-m^2e^{2\Phi(r)}),\,\,\,\,\,\,\,\,\,g(r)^2=1-\frac{b(r)}{r}.
\end{equation}
If we take 
\begin{equation}
e^{r}=\frac{f}{g}dr,\,\,\,\,\,\,\,\,\,e^{\phi}=fr d\phi,
\end{equation}
and using the Cartan structural equation we have
\begin{eqnarray}
\omega^{\phi}_{\,\,\,\,r}=(fr)' \frac{g}{f^2 r} d\phi\,\,\,\,\, \Rightarrow \,\,\,\,\, d\omega^{\phi}_{\,\,\,\,r}=\left((fr)''\frac{g}{f}+(fr)'\left(\frac{g}{f}\right)'\right)\frac{g}{f^2 r}dr\wedge d\phi,
\end{eqnarray}
hence, the Gaussian curvature  is
\begin{equation}\label{gaussc}
K_{G}=-\left((fr)''\frac{g}{f}+(fr)'\left(\frac{g}{f}\right)'\right)\frac{g}{f^2 r}\,.
\end{equation}

Replacing (\ref{frgr}) in (\ref{gaussc}) and taking $\Phi(r)=0$ we obtain:
\begin{equation}\label{gaussJacobi}
K_{G}=\frac{\left(r b'-b\right)}{2 r^3(E^2-m^2)}\,.
\end{equation}

When the condition $E^2>m^2$ is satisfied\footnote{On the other hand, when $E=m$ we can see that the Jacobi metric vanishes, therefore our approach does not work anymore. For the case $E^2<m^2$ we can see that the Gauss curvature changes its sign and therefore the type of trajectories also does change. The Gaussian curvature diverges at $E=m$ and then it changes its sign, the method does not allow us to know what is happening at the point where the Gauss curvature goes to infinity.}, the factor that determines the sign of the Gauss curvature is $\left(b'r-b\right)$, which turns out to be the relevant part of the flare-out condition\footnote{The flare-out condition $\left(b'r-b\right)<0$ has to be satisfied in order to have a  minimal radius at the throat. It means that the null rays through the throat are divergent, which in its turn implies the violation of the null energy condition} of the wormhole. Due to the fact that the Gaussian curvature is an intrinsic quantity independent of the coordinate system  we claim that we have found a direct relationship between the Gaussian curvature of the Jacobi metric with the flare-out condition. The coordinate dependent flare-out condition $b'r-b<0$ can be generalized to a covariant definition \cite{Hochberg:1997wp} that depends on the extrinsic curvature, and therefore, we have made the first step towards the study of the projection of the covariant flare-out condition over the Jacobi metric. We have provided evidence for the fact that only with the information contained in the Jacobi metric we were able to determine the existence of a throat. Therefore the information of the covariant flare-out condition has to be contained in the Jacobi metric.

The negativity of the Gauss curvature implies that the only possibility for the Jacobi metric is to have a negative intrinsic curvature. It is related with the fact that there is only one stable circular orbit at the throat: all throat points are saddle points. See section \ref{stableorbit}. Indeed, being the intrinsic geometry such that the Gauss curvature is negative it is expected that the only stable circular orbit is at the saddle points, namely the throat. It implies that we only need to know the sign of the Gauss curvature in order to determine the existence of a throat, and therefore a wormhole. In fig. \ref{closed:1} we plotted the solution (\ref{firstordeq:2}) for the size of the throat $bo=2$ and $bo=3$. We can clearly see that for $C=\frac{1}{bo^2}$ the points at the throat are saddle points, and therefore we will have a stable circular orbit. \\
When $E^2<m^2$ we have that the curvature is positive. We know that there still exist geodesics but our approach does not tell anything particular. The geodesics will correspond to the case when $C<0$ whose trajectories are given by (\ref{orbitsCn}). 
The information about geodesics is all encoded in the Jacobi metric. \\
In \cite{Hochberg:1997wp} the flare-out condition was generalized, they define the throat as a two-dimensional minimal surface taken in any constant-time slice, then three types of  flare-out conditions are defined using normal Gaussian coordinates\footnote{Taking $\overrightarrow{n}$ the vector normal to the surface. $K$ is the trace of the extrinsic curvature, it will depend on how the surface is embedded in the ambient space. The authors make three definitions of flare-out conditions, the strong condition which requires that at a point $x$ the trace satisfies $K=0$ and everywhere $\frac{\partial K}{\partial \overrightarrow{n}}\leq 0$. Having the strict inequality at the point $x$.  The weak condition is satisfied if $K=0$ and $\int \sqrt{g}\frac{\partial K}{\partial  \overrightarrow{n}}d^2 x<0$. Finally, the simple flare-out condition, which requires $K=0$ and $\frac{\partial K}{\partial \overrightarrow{n}}\leq 0$ everywhere.}.\\
The flare-out condition is coordinate dependent, what we have found has to be taken in that context. We should study a covariant formulation in order to get a better insight \cite{Arganaraz:2020}. \\
In our study, equation (\ref{gaussJacobi}) shows a relationship between an intrinsic quantity, such as the Gauss curvature, and the flare-out condition, which in general grounds is related to the surface embedding in a higher dimensional space \footnote{It is important to note that we are calculating the gauss curvature of the Jacobi metric, corresponding to (\ref{restrictedm:1}), which is a two dimensional metric (because we are taking surfaces of constant energy, and because of spherical symmetry) and therefore the trace of the Ricci tensor is the Gaussian curvature.}. 

\subsection{Exotic matter}
The Einstein equations restrict the matter that can sustain a wormhole. In wormhole theory, we take the ans\"{a}tze for the metric and use it for calculating the geometric quantities in one side of the equation, then using the fact that the Einstein tensor should have the same structure of the stress-energy tensor, we find the form that the matter needs to have in order to satisfy Einstein equations. For the case of the metric (\ref{wormhole}), and taking $\Phi(r)=0$, the diagonal components of the stress energy tensor in an orthonormal basis vectors\footnote{Here we use orthonormal vectors $(\textbf{e}_{\hat{t}}=e^{-\Phi}\textbf{e}_{t},\textbf{e}_{\hat{r}}=(1-b/r)^{(1/2)}\textbf{e}_{r},\textbf{e}_{\hat{\theta}}=(1-b/r)^{(1/2)}\textbf{e}_{\theta},\textbf{e}_{\hat{\phi}}=(r\sin(\theta))^{-1}\textbf{e}_{\phi}, )$, where $(\textbf{e}_{t},\textbf{e}_{r},\textbf{e}_{\theta},\textbf{e}_{\phi})$ are the basis vectors in the coordinate system. }, are
\begin{eqnarray}
T_{\hat{t}\hat{t}}=\rho(r)&=&\frac{1}{8\pi}\frac{b'}{r^2}\label{stressenergy1}\\
-T_{\hat{r}\hat{r}}=\tau(r)&=&\frac{1}{8\pi}\frac{b}{r^3}\label{stressenergy2}\\
T_{\hat{\theta}\hat{\theta}}=T_{\hat{\phi}\hat{\phi}}=p(r)&=& -\frac{1}{8\pi}\left(1-\frac{b}{r}\right)\left(\frac{b'r-b}{2r^3(1-b/r)}\right)\label{stressenergy3}
\end{eqnarray}
The Gauss curvature in equation (\ref{gaussJacobi}) can be rewritten using (\ref{stressenergy1}) and (\ref{stressenergy2}):
\begin{equation}
K_{G}=\frac{\pi(\rho-\tau)}{E^2-m^2}.
\end{equation}

The negativity of $K_{G}$ is directly related with the fact that the only possible way to sustain a wormhole, in general relativity, is by accepting the existence of exotic matter\footnote{The exotic matter violates all energy conditions and it is considered unphysical.} at the throat.  Thus, for a fixed energy such that $E^2>m^2$ the negativeness of the Gauss curvature is tantamount to the existence of a flare-out condition and therefore to the existence of exotic matter.

In the following section we discuss about the Kepler problem. There should be a relationship between the flare-out condition, the type of trajectories and the matter sustaining the wormhole.

\subsection{On the Kepler problem}
One of the advantages of the Jacobi metric formalism is that it allows the geometrization of the dynamics. In \cite{Pin:1975} the author shows that the trajectories for the Kepler problem in Minkowski space can be classified by the sign of the Gaussian curvature of the Jacobi metric. The evidence presented in section \ref{flareout} lead us to conclude that it can be done in a Morris-Thorne wormhole background. We are going to see what happens in a very particular case of the Kepler problem for a static spherically symmetric wormhole. In order to do that we consider a radial potential $U(r)=-\frac{\alpha}{r}$. Because spherical symmetry we are entitled to work with the spatial part of the metric, and because of angular momentum conservation we only consider two dimensions. The Jacobi metric is given by\footnote{In general, the potential function $U$ will depend on $r'\neq r$ where $r'$ is the distance between both particles. The particular metric (\ref{Keplermetric}) will represent a particle moving in a central potential where the other particle is located in the origin of coordinates. Moreover, we will have the restriction $r>r_{o}$, meaning that the particle cannot go beyond the throat.} \cite{Chanda:2016sjg}
\begin{equation}\label{Keplermetric}
ds^2_{K}=2(E-U(r))\left(\frac{dr^2}{1-\frac{b(r)}{r}}+r^2d\phi^2\right),
\end{equation}
and, as before, we can write the previous metric in the form   (\ref{metricfg}) with
\begin{equation}
f(r)^2=2(E-U(r)),\,\,\,\,\,\,g(r)^2=1-\frac{b(r)}{r}\,.
\end{equation}
Then, using equation (\ref{gaussc}) and taking $\Phi=0$ we can find a general expression for the Gaussian curvature
\begin{equation}\label{gaussk1}
K_{G}=\frac{(4E^2r^2+2\alpha^2+6Er\alpha)(rb'-b)- 4Er\alpha(r-b)}{16r^2(E r+\alpha)^3}
\end{equation}
When $E=0$ the equation (\ref{gaussk1}) becomes
\begin{equation}
K_{G}=\frac{rb'-b}{8\alpha r^2},
\end{equation}
then, because of the flare-out condition, we conclude that  $K_G<0$. Therefore, all trajectories are hyperbolic. Similarly, when $E>0$ we have that all trajectories are hyperbolic. However, when $E<0$ we cannot known what is happening with the sign. The equation (\ref{gaussk1}) cannot be written as a function of the flare-out condition only. At the throat we have that $r_{o}=b(r_{o})$ and therefore $K_{G}<0$,which implies that the trajectories at the throat are elliptic (closed). In particular, when $b(r)$ is given by (\ref{shapef}) with $n=2$ then, because of the flare-out condition \footnote{The wormholes with shape function $b(r)=b_{o}^n r^{1-n}$ do satisfy the flare-out condition. Indeed $r b'(r)-b=-\frac{2b_{o}^2}{r}$}, the Gauss curvature is positive provided that $\vert {b_{o}}\vert<w(r)$ where
\begin{equation}
w(r)=\sqrt{\frac{-Er^3\alpha}{(Er+\alpha)^2+E^2r^2}},
\end{equation}
which in its turn implies that the trajectories are going to be elliptic. 
Thus, 

\begin{equation}
E>-\frac{\alpha}{r}\,\,\,\,\left\lbrace  
\begin{array}{ll}
E<0,\vert b_{o}\vert <w(r)\Rightarrow K_{G}>0 \Rightarrow\,\, elliptic\\
E=0 \Rightarrow  K_G<0 \Rightarrow hyperbolic \\
E>0 \Rightarrow  K_G<0 \Rightarrow hyperbolic
\end{array}
\right. 
\end{equation}

We will have closed periodic orbits described by (\ref{Keplermetric}) when $-\frac{\alpha}{r}<E<0$, and $\vert b_{o}\vert <w(r)$. For a given $\alpha$ and $E$, the $w(r)$ function is an increasing function of $r$ and therefore $b_{o}$ can be chosen in order to satisfy the required inequality\footnote{As $r$ increases the range of possible values for $b_{o}$ increases.}. This one of our major results. As we have seen, the Kepler problem can be solved in this kind of static spaces. Moreover, the same analysis applies for dynamic wormholes.

\subsection{Matter and trajectories}

The curvature (\ref{gaussk1}) can be rewritten using (\ref{stressenergy1}),(\ref{stressenergy2}) and (\ref{stressenergy3}). We can classify all trajectories for the Kepler problem using the matter terms of the Einstein equations. Thus,
\begin{equation}
K_{G}=\frac{4\pi r(2E^2r^2+\alpha^2+3Er\alpha)(\rho-\tau)-E\alpha(1-8\pi r^3 \tau)}{4(Er+\alpha)^3}
\end{equation}
Then, using the fact that $\rho-\tau <0$ and the shape function (\ref{shapef}) we can classify the trajectories
\begin{equation}
E>-\frac{\alpha}{r}\,\,\,\,\left\lbrace  
\begin{array}{ll}
E<0, \tau>\frac{-E \alpha+4 \pi r(2E^2r^2+Er\alpha+\alpha^2) \rho}{4 \pi r(2E^2r^2+Er\alpha+\alpha^2)}  \Rightarrow K_{G}>0 \Rightarrow\,\, elliptic\\
E=0 \Rightarrow  K_G<0 \Rightarrow hyperbolic \\
E>0 ,  \tau< \frac{1}{8\pi r^3} \Rightarrow  K_G<0 \Rightarrow hyperbolic
\end{array}
\right. 
\end{equation}
This analysis lead us to think about a very important issue, namely the possible theories(modified gravities) that admit certain trajectories for the Kepler problem. Due to the fact that the flare-out condition implies the violation of the NEC (Null energy condition), the authors in \cite{Harko:2013yb} found a zone where the NEC is violated, thus in the case of $R^2$ gravity \footnote{$R^2$ gravity is the a particular case of an $f(R)$ gravity, where $f(R)=R+\frac{1}{2}\alpha R^2$ } and in the throat they found, for the Morris-Thorne wormhole
\begin{equation}\label{NEC}
\frac{T_{\mu\nu}k^{\mu}k^{\nu}\mid_{ r_{o}}}{1+\alpha R}<\frac{\alpha (1-b')(2b'-rb'')}{k^2r^4(r^2+2\alpha b')}\mid_{ r_{o}}.
\end{equation} 
Certainly, the flare-out condition implies the violation of the NEC  at the throat, however there are ways to circumvent this problematic issue. They introduce an effective stress-energy tensor made of a sum of the matter tensor plus the geometric terms corresponding to the higher order terms. It is possible to find an $\alpha$ such that (\ref{NEC}) is satisfied even in the case that $T_{\mu\nu}k^{\mu}k^{\nu}\geq 0$.

It will be interesting to study what kind of modified gravities admit closed trajectories for the Kepler problem.

\section{Discussion and final remarks}\label{discussion}
We have studied the dynamics of wormholes. Instead of studying it by the usual techniques, namely  solving the geodesic equation, we have used the Jacobi metric approach. This approach helped us to study trajectories in a Lorentzian manifold as if it where geodesics of a Riemannian manifold where the Jacobi metric is defined. In particular, the study of wormholes was eye-opening, specially regarding the properties of the spacetime at the throat of the wormhole. We calculated the Jacobi metric and used it for finding the first integral of the trajectory equations. In the case of the Morris-Thorne wormhole we have found that the only stable circular orbit is at the throat. This fact is related with another interesting situation; the Gauss curvature of the Jacobi metric at the throat is negative and therefore the closed trajectories, if any,  are at the throat only, where all the points are saddle points. Moreover, using Jacobi elliptic functions we were able to find a simple equation for describing the orbits. We have shown that the flare-out condition is in direct relationship with the Gaussian curvature of the Jacobi metric restricting the sign of this curvature. This is something remarkable, we only need to calculate the Gaussian curvature of the Jacobi metric in order to know if a spacetime has the characteristics of a wormhole, in other words, the existence of a throat can be inferred by using information stored in the Jacobi metric only. Take as an example the exponential metric. In \cite{Boonserm:2018orb} it is shown that the exponential metric is a wormhole. Now, consider the Jacobi metric corresponding to the exponential metric (\ref{jacobiexp}). It can be shown that the Gaussian curvature of this metric is negative, therefore the metric (\ref{expm}) has to have a collection of saddle points and therefore a throat. It remains to see if the method works for other than asymptotically spherically symmetric spacetimes . We have also studied a restriction of the Kepler problem in the Morris-Thorne wormhole and we were able to classify all the trajectories by using the sign of the Gauss curvature. Using the fact that the flare-out condition for a wormhole, satisfying Einstein equations, can be written as a function of the diagonal components of the stress-energy tensor, we were able to classify the trajectories using the properties of the matter involved.
An interesting problem would be to study the kind of gravity theories that allow the existence of closed trajectories in a wormhole spacetime. We have to remember that the flare-out condition can be related with the matter sustaining the wormhole, and therefore the Gaussian curvature can be related too. However, we know that the  flare-out condition is coordinate dependent, therefore we need to make a covariant approach. We know that a restricted variational principle in an arbitrary metric and constant particle energy is completely equivalent to a variational principle, which now is unrestricted, but defined in the Jacobi metric, which involves the particle energy in an explicit way. Therefore, the paths of a given constant energy are still geodesics of the Jacobi metric. On this context, and knowing that the geodesics equation can be obtained using the classical Hamilton-Jacobi method with the Jacobi metric, the Jacobi approach has similar limitations as the classical method, although it has more advantages in other aspects. One prominent example is the solution of the Kepler problem. We have studied the problem in our wormhole metric, but we have discovered, again, that the flare-out condition determines the sign, and therefore, the type of geometries allowed. 

The dynamics of massive particles in wormhole spacetimes, specially in the Morris-Thorne wormhole, has been studied in the literature. The Jacobi metric approach has new advantages. It provides a simpler way to find the equation \eqref{radialeq:1}, which is the first integral of motion. This equation depends only on the redshift function $\Phi(r)$ and the shape function $b(r)$. In the simplest case $\Phi=0$ we where able to find analytically all trajectories. The only closed trajectory is an stable circular orbit, and it is located at the throat. This result applies for the general shape form \eqref{shapef2}, and therefore it works for any $n$. In figure \eqref{closed:1} we have plotted the solution \eqref{sol} for a trajectory satisfying \eqref{cfinal} and for two different throat sizes. In figure \eqref{open:2} we have plotted the open trajectories. In all figures, the trajectories cannot go beyond the throat. This particular behaviour lead us to think about the characterization of the wormhole geometries. The existence of only one stable circular orbit at the throat shows that this zone is special and could be characterized by a purely geometric intrinsic quantity. Later on we discovered that we can use, for this purpose, the Gaussian curvature.\\
The procedure can be generalized for wormholes with $\Phi \neq 0$, and for stationary wormholes on which we have to use the so called Eisenhart-Duval lift. See appendix \ref{dynamicwormhole} for details. Finally, it will be useful to extend the study stationary wormholes, in a similar way that stationary black hole metrics have been studied in \cite{Chanda:2019guf}.\\
 
 
\section{Appendix}\label{appendix}

\subsection{Jacobi metric for different wormhole metrics}
\subsubsection{The exponential metric}
An interesting example of a wormhole is the exponential metric. This metric was known to be horizonless, however, it was not until the reference \cite{Boonserm:2018orb} appeared that it started to be considered as a wormhole. This metric has nice features, for example, it is a traversable wormhole, with time slowed down on  the other side of throat. The innermost stable  circular orbits and unstable photon orbits exist and  are  a little bit  shifted  from where they would be located in Schwarzschild spacetime. For more details see \cite{Boonserm:2018orb}.
The exponential metric is a wormhole given by
\begin{equation}\label{expm}
ds^{2}=-e^{-\frac{2m}{r}}dt^2+e^\frac{2m}{r}\{dr^{2}+r^{2}(d\theta^{2}+\sin^{2}(\theta)d\phi^{2})\}.
\end{equation}
Using the same procedures as in previous sections we can calculate the Jacobi metric: 
\begin{equation}
ds^2=\frac{E^{2}-m^{2}e^{-\frac{2m}{r}}}{e^{-\frac{2m}{r}}}e^{\frac{2m}{r}}\{dr^{2}+r^{2}(d\theta^{2}+\sin^{2}{\theta}d\phi^{2})\}.
\end{equation}
 Taking $\theta=\frac{\pi}{2}$ we have
 \begin{equation}\label{jacobiexp}
ds^2=\frac{E^{2}-m^{2}e^{-\frac{2m}{r}}}{e^{-\frac{2m}{r}}}e^{\frac{2m}{r}}(dr^{2}+r^{2}d\phi^{2}).
\end{equation}
The Clairut constant is
\begin{equation}\label{lexp}
l=\left(E^{2}-m^{2}e^{-\frac{2m}{r}}\right)e^{\frac{4m}{r}}r^{2}\left(\frac{d\phi}{ds}\right)=cte.
\end{equation} 

Using (\ref{lexp}) in (\ref{jacobiexp}) we get
\begin{equation}
\left(E^{2}-m^{2}e^{-\frac{2m}{r}}\right)^{2}e^{\frac{4m}{r}}\left(\frac{dr}{ds}\right)^{2}=E^{2}-m^{2}e^{-\frac{2m}{r}}-\frac{l^{2}}{r^{2}}.
\end{equation}
We recover the usual expression \cite{Boonserm:2018orb} 
\begin{equation}
m^{2}e^\frac{4m}{r}\left(\frac{dr}{d\tau}\right)^{2}=E^{2}-m^{2}e^{-\frac{2m}{r}}-\frac{l^{2}}{r^{2}},
\end{equation}

provided that 
\begin{equation}
d\tau=\frac{m e^\frac{2m}{r}}{E^{2}-m^{2}e^{-\frac{2m}{r}}}ds.
\end{equation}

As we can easily see from the previous calculations we can proceed with a similar study as the one carried in this paper but for the exponential metric or for almost any other wormhole metric.

\subsubsection{Darmour-Solodukin wormholes}
The Darmour-Solodukin wormhole is a modification of the Schwarzschild metric in order to make it horizonless. Usually thought as a good candidate for cosmological observations, the "black hole foils" are objects that mimic some aspects of black holes, while lacking some of their defining features, such as the horizon \cite{Damour:2007ap}.
The Darmour-Solodukin wormhole metric is  
\begin{equation}
ds^{2}=-\left(1+\frac{2GM}{r}+\lambda^{2}\right)dt^{2}+\frac{dr^2}{\left(1-\frac{2GM}{r}\right)}+r^{2}(d\theta^{2}+\sin^{2}(\theta)d\phi^{2}).
\end{equation}
The Jacobi metric is then
\begin{equation}
ds^2=\frac{E^2-m^2\left(1-\frac{2GM}{r}+\lambda\right)}{\left(1-\frac{2GM}{r}+\lambda^2 \right)}\left(\frac{dr^2}{\left(1-\frac{2GM}{r}\right)}+r^2d\phi^2 \right),
\end{equation}

where we have set $\theta=\frac{\pi}{2}$. Then, the Clairut constant is
\begin{equation}
l=\frac{\left(E^2-m^2\left(1-\frac{2GM}{r}+\lambda^2 \right)\right)r^2}{\left(1-\frac{2GM}{r}+\lambda^2 \right)}\left(\frac{d\phi}{ds}\right).
\end{equation}

Hence

\begin{equation}
\frac{E^2-m^2\left(1-\frac{2GM}{r}+\lambda^2 \right)}{\left(1-\frac{2GM}{r}+\lambda^2 \right)\left(1-\frac{2GM}{r}\right)}\left(\frac{d r}{ds}\right)^2=E^2-\left(1-\frac{2GM}{r}+\lambda^2\right)\left(m+\frac{l^2}{r^2}\right),
\end{equation}

if we make 
\begin{equation}
d\tau =\frac{m\left(1-\frac{2GM}{r}+\lambda^2 \right)}{E^2-m^2\left(1-\frac{2GM}{r}+\lambda^2 \right)}ds.
\end{equation}

We are ready for studying charged wormholes. The dynamics is going to be more complicated that their uncharged counterparts , but we can apply the same procedure without any problem.

\subsubsection{The scalar charged wormhole}

Charged wormholes can be found by introducing matter to an static wormhole which is already sustained by exotic matter. The  charges  play  the  role  of  the  additional  matter \footnote{We neglect the interaction between the matter and the scalar field. In addition, we consider that the scalar field does not radiate}. In \cite{Kim:2001ri} the author introduced charged wormholes. In particular he works with scalar charged wormholes, whose metric is given by
\begin{equation}\label{mcharged}
ds^{2}=-dt^{2}+\frac{dr^{2}}{\left(1-\frac{b(r)}{r}+\frac{\eta}{r^{2}}\right)}+r^{2}(d\theta^{2}+\sin^{2}(\theta)d\phi^{2}),
\end{equation}

where $\eta$ is the scalar charge associated to the scalar field, and  the shape function $b(r)$
 is restricted to be 
 \begin{equation}
b=b_{o}^\frac{2\beta}{2\beta+1}r^\frac{1}{2\beta+1}\,\,\,\,\,\beta<1/2.
\end{equation}
The Jacobi metric is 
\begin{equation}\label{jcharged}
ds^2=(E^2-m^2)\left(\frac{dr^2}{\left(1-\frac{b(r)}{r}+\frac{\eta}{r^2}\right)}+r^2d\phi^2\right).
\end{equation} 
where we have set $\theta=\frac{\pi}{2}$.
The Clairut constant is 
\begin{equation}\label{ccharged}
l=(E^2-m^2)r^2\left(\frac{d\phi}{ds}\right).
\end{equation}
Using (\ref{jcharged}) and (\ref{ccharged}) we found
\begin{equation}
\frac{(E^2-m^2)^2}{\left(1-\frac{b}{r}+\frac{\eta}{r^2}\right)}\left(\frac{dr}{ds}\right)^2+\frac{l^2}{r^2}=E^2-m^2.
\end{equation}

Setting 

\begin{equation}
d\tau=\frac{m}{E^2-m^2}ds,
\end{equation}

we recover the expected results
\begin{equation}
\frac{m^2}{\left(1-\frac{b}{r}+\frac{\eta}{r^2}\right)}\left(\frac{dr}{d\tau}\right)^2=E^2-m^2-\frac{l^2}{r^2}.
\end{equation}

In the same article \cite{Kim:2001ri} the author introduced electrically charged wormholes, in the next section we present  the most general case, the modified charged wormhole.
\subsubsection{Modified charged wormhole}
There is a way to generalize even more the charged wormhole. The new generalization is called modified charged wormhole \cite{Kuhfittig:2011xh} and its metric is given by

\begin{equation}
ds^{2}=-\left(1+R(r)+\frac{Q^{2}}{r^2}\right)dt^{2}+\frac{dr^{2}}{\left(1-\frac{b(r)}{r}+\frac{Q^{2}}{r^{2}}\right)}+r^{2}(d\theta^{2}+\sin^{2}(\theta)d\phi^{2}),
\end{equation}
where $R(r)$ is any positive function of $r$, and when $R(r)=0$ we recover the electrically charged wormhole given in \cite{Kim:2001ri}.

The Jacobi metric for the modified charged wormholes is
\begin{equation}\label{mcjacobi}
J_{ij}=\frac{E^2-m^2\left(1+R(r)+\frac{Q^2}{r^2}\right)}{\left(1+R(r)+\frac{Q^2}{r^2}\right)}\left(\frac{dr^{2}}{\left(1-\frac{b(r)}{r}+\frac{Q^{2}}{r^{2}}\right)}+r^{2}d\theta^{2}+d\phi^{2}\right).
\end{equation}

The Clairut constant is
\begin{equation}\label{mcclairut}
l=\frac{E^2-m^2\left(1+R(r)+\frac{Q^2}{r^2}\right)r^2}{\left(1+R(r)+\frac{Q^2}{r^2}\right)}\left(\frac{d\phi}{ds}\right).
\end{equation}

Then, using (\ref{mcjacobi}) and (\ref{mcclairut}) we obtain 
\begin{equation}
\frac{\left(E^2-m^2\left(1+R(r)+\frac{Q^2}{r^2}\right)\right)^2}{\left(1+R(r)+\frac{Q^2}{r^2}\right)^2\left(1-\frac{b}{r}+\frac{Q^2}{r^2}\right)}\left(\frac{dr}{ds}\right)^2=\frac{E^2}{\left(1+r+\frac{Q^2}{r^2}\right)}-m^2-\frac{l^2}{r^2}.
\end{equation}

which after setting 
\begin{equation}
d\tau=\frac{m\left(1+R(r)+\frac{Q^2}{r^2}\right)}{E^2-m^2\left(1+R(r)+\frac{Q^2}{r^2}\right)}ds.
\end{equation}

becomes

\begin{equation}
\frac{m^2}{\left(1-\frac{b}{r}+\frac{Q^2}{r^2}\right)}\left(\frac{dr}{d\tau}\right)^2=\frac{E^2}{\left(1+R+\frac{Q^2}{r^2}\right)}-m^2-\frac{l^2}{r^2}.
\end{equation}

\subsection{Another redshift function}\label{anotherf}

As we have seen, we can find the Jacobi metric straightforwardly. Depending on the redshift function the calculation is going to increase in difficulty. Here we present a brief example with a non-zero redshift function.

We take the redshift function $e^{2\Phi}=\left(1-b(u)u\right)+\epsilon(u)$, where $\epsilon$ is a continuous function with vanishing contribution far from the throat. The respective Jacobi metric, after setting $\theta=\frac{\pi}{2}$,  is written as
\begin{equation}
ds^2=\frac{E^2-m^2(1-\frac{b(r)}{r}+\epsilon(r))}{1-\frac{b(r)}{r}+\epsilon(r)}\left(\frac{dr^2}{1-\frac{b(r)}{r}}+r^2d\phi^2\right).
\end{equation}
In this particular case the Clairut constant is given by
\begin{equation}
l=\frac{E^2-m^2(1-\frac{b(r)}{r}+\epsilon(r))}{1-\frac{b(r)}{r}+\epsilon(r)}r^2\left(\frac{d\phi}{ds}\right),
\end{equation}

hence
\begin{equation}
\frac{E^2-m^2(1-\frac{b(r)}{r}+\epsilon(r))^2}{(1-\frac{b(r)}{r}+\epsilon(r))^2\left(1-\frac{b}{r}\right)}\left(\frac{dr}{ds}\right)^2+\frac{l^2}{r^2}=\frac{E^2}{1-\frac{b(r)}{r}+\epsilon(r)}-m^2.
\end{equation}

Taking
\begin{equation}
d\tau=\frac{m\left(1-\frac{b(r)}{r}+\epsilon(r)\right)}{E^2-m^2\left(1-\frac{b(r)}{r}+\epsilon(r)\right)}ds,
\end{equation} 

we recover (see eq.(44) in \cite{Mishra:2017yrh})
\begin{equation}
\frac{m^2}{1-\frac{b(r)}{r}}\left(\frac{dr}{d\tau}\right)^2=\frac{E^2}{1-\frac{b(r)}{r}+\epsilon(r)}-m^2-\frac{l^2}{r^2}.
\end{equation}

Then, we can start the analysis of bounded and unbounded orbits.For the classical analysis we refer the reader to \cite{Mishra:2017yrh}. Instead of doing that, we are going to study dynamical wormholes. In this particular case we have to modify the metric in order to use a Jacobi metric approach.

\subsection{Dynamic wormhole geometries: the Eisenhart-Duval lift}\label{dynamicwormhole}
When considering a dynamic wormhole some of the coefficients of the wormhole metric will depend on time, and therefore we cannot use the Jacobi formalism directly. We have to adapt the technique for finding the Jacobi metric. Usually, in time dependent systems, the energy is not conserved, an therefore the systems become dissipative. The usual Jacobi method \footnote{We call usual method to the time-independent one.}
projects  the geodesics to a constant energy hypersurface. When the metric is dynamic we do not have a surface where we can project. However, there is a way to circumvent this problem, we can use the Eisenhart-Duval lift\cite{Chanda:2016aph}. This lift help us to embed non-relativistic theories into a Lorentzian geometry. This method allows us to geometrize classical systems by building a dynamically equivalent system but in higher dimensional Lorentzian space. For a good introduction see \cite{Cariglia:2015bla,Finn:2018cfs}.

Let us consider the following time dependent wormhole metric
\begin{equation}\label{dynamicw}
ds^{2}=-c^2e^{2\Phi}dt^{2}+a^{2}(t)\left[\frac{dr^{2}}{\left(1-\frac{b(r)}{r}\right)}+r^{2}d\theta^{2}+r^{2}\sin^{2}(\theta)d\phi^{2}\right].
\end{equation}
As before $e^{2\Phi}$ is the redshift function and $b(r)$ is the shape form. The function $a(t)$ is a continuous function of time. We modify the metric (\ref{dynamicw}) by introducing a dummy variable $\sigma$ in the following way
\begin{equation}
ds^{2}=-c^2e^{2\Phi}dt^{2}+a^{2}(t)\left[\frac{dr^{2}}{\left(1-\frac{b(r)}{r}\right)}+r^{2}d\theta^{2}+r^{2}\sin^{2}(\theta)d\phi^{2}\right]+2cdtd\sigma.
\end{equation}
The corresponding line element Lagrangian becomes
\begin{equation}\label{lag}
\mathcal{L}(r,\dot{r},t)=\frac{m}{2}\left(-c^2V^{2}(r,t)\dot{t}^{2}+2c\dot{\sigma}\dot{t}-g_{ij}(r,t)\dot{x}^{i}\dot{x}^{j}\right),
\end{equation}
where $V^{2}(r,t)=-e^{2\Phi}$.
The Jacobi metric is 
\begin{equation}
ds^2=(2qp_{t}-q^2c^2V^2(r,t)-m^2c^2)a^{2}(t)\left(\frac{dr^{2}}{\left(1-\frac{b(r)}{r}\right)}+r^{2}d\theta^{2}+r^{2}\sin^{2}(\theta)d\phi^{2}\right).
\end{equation}
where $q$ is a conserved quantity and $p_{\sigma}=qc=cte.$

Taking $\theta=\frac{\pi}{2}$ we got
\begin{equation}
ds^2=(2qp_{t}-q^2c^2V^2(r,t)-m^2c^2)a^{2}(t)\left(\frac{dr^{2}}{\left(1-\frac{b(r)}{r}\right)}+r^{2}d\phi^{2}\right).
\end{equation}
Hence,
\begin{equation}\label{eqdynamic}
1=(2qp_{t}-q^2c^2V^2(r,t)-m^2c^2)a^{2}(t)\left(\frac{1}{\left(1-\frac{b(r)}{r}\right)}\left(\frac{dr}{ds}\right)^2+r^{2}\left(\frac{d\phi}{ds}\right)^2\right).
\end{equation}
In this case the Clairaut constant is 
\begin{equation}\label{cldynamic}
l=(2qp_{t}-q^2c^2V^2(r,t)-m^2c^2)a^2(t)r^2\left(\frac{d\phi}{ds}\right)^2.
\end{equation}
Using (\ref{cldynamic}) and (\ref{eqdynamic}) we find
\begin{equation}\label{eq2dyn}
(2qp_{t}-q^2c^2V^2(r,t)-m^2c^2)\left(\frac{dr}{ds}\right)^2=\frac{1}{a^2(t)}\left(1-\frac{b(r)}{r}\right)\left(2qp_{t}-q^2c^2V^2(r,t)-m^2c^2-\frac{l^2}{a^2(t)r^2}\right).
\end{equation}
Then, with the change of parameter
\begin{equation}
ds=\frac{1}{m}(2qp_{t}-q^2c^2V^2(r,t)-m^2c^2)d\tau,
\end{equation}
 we obtain 
 \begin{equation}
 m^2\left(\frac{dr}{d\tau}\right)^2=\frac{1}{a^2(t)}\left(1-\frac{b(r)}{r}\right)\left(2qp_{t}-q^2c^2V^2(r,t)-m^2c^2-\frac{l^2}{a^2(t)r^2}\right).
 \end{equation}
We can see that $p_{\sigma}=qc$ by taking the derivative with respect to $\dot{\sigma}$ in the Lagrangian (\ref{lag}). Similarly we can find
\begin{equation}
p_t=c^2e^{2\Phi(r)}q-\frac{mc\dot{\sigma}}{2}.
\end{equation} 
Therefore, equation (\ref{eq2dyn}) becomes
\begin{equation}
m^2\left(\frac{dr}{d\tau}\right)^2=\frac{1}{a^2(t)}\left(1-\frac{b(r)}{r}\right)\left(e^{2\Phi(r)}q^2c^2-mcq^2\dot{\sigma}-m^2c^2-\frac{l^2}{a^2(t)r^2}\right).
\end{equation}
The previous equation will give as a result geodesics in 5 dimensions, we project to four dimensions by taking\footnote{It will imply that $p_{\sigma}=-cm \dot{t}\neq cte $.} $\dot{\sigma}=0$, then
\begin{equation}
\left(\frac{dr}{d\tau}\right)^2=\frac{1}{a^2(t)}\left(1-\frac{b(r)}{r}\right)\left(e^{2\Phi}q^2c^2-c^2-\frac{l^2}{m^2a^2(t)r^2}\right).
\end{equation}

We can see that the dynamic wormhole have to be treated differently. Moreover, there are much more subtleties  when dealing with this spacetimes.

\subsection{The classical approach: Weierstrass functions}\label{eqsol} 
The eq. (\ref{firstordeq:2}) can be solved by using Weierstrass elliptic functions\footnote{The Weierstrass elliptic function $\mathcal{P}(\phi)$ is defined as the inverse of $\phi=\int_{y}^{\infty}\frac{dt}{\sqrt{4t^3-g_{2}t-g_{3}}}$. Where $g2$ and $g3$ are constants. Therefore $y=\mathcal{P}(\phi)$} $\mathcal{P}(\phi)$. The right hand side has at least a real root $\alpha$, then with the change of variable $s=u-\tilde{\beta}$ the eq (\ref{firstordeq:2}) transforms to
\begin{equation}\label{eqtrans:1}
(s')^{2}=Ps^{4}+Qs^{3}+Rs^{2}+Ts,
\end{equation}
with
\begin{eqnarray}
P&=&b_{o}^{2}\\
Q&=&4b_{o}^{2}\tilde{\beta}\\
R&=&b_{o}^{2}\left(6\tilde{\beta}^{2}-\left(C+\frac{1}{b_{o}^{2}}\right)\right)\\
T&=& b_{o}^{2}\left(4\tilde{\beta}^{3}-2\left(C+\frac{2\tilde{\beta}}{b_{o}^{2}}\right)\right)
\end{eqnarray}
We make the substitution $\psi=\frac{4y}{T}-\frac{R}{3T}$ into the eq. (\ref{eqtrans:1}) and we get
\begin{equation}
(y')^{2}=4y^{3}-g_{2}y-g_{3},
\end{equation}
where
\begin{eqnarray}
g_{2}&=&-\frac{1}{4}\left(QT-\frac{2}{3}R^{3}+\frac{1}{3}R^{2}\right)\\
g_{3}&=&-PT^{2}-\frac{2R^{3}}{27}
\end{eqnarray}
then, the solution for (\ref{radialeq:1}) is
\begin{equation}
\frac{1}{r(\phi)}=\beta+\frac{T}{4 \mathcal{P}(\phi+\zeta_{o})},
\end{equation}
where $\mathcal{P}$ represents the Weierstrass elliptic function, and $\zeta_{o}$ is a complex parameter. A similar result, with another approach was found in \cite{Muller:2008zza}.\\
The orbits for massive particles are given by
\begin{equation}
\frac{1}{r}=-\sqrt{C}+ b_{o}^{2}\left(5C+\frac{1}{b_{o}^{2}}\right)\frac{1}{4 \mathcal{P}(\phi+\zeta_{o})}.
\end{equation}

\break


\begin{thebibliography}{9}
\bibitem{Morris:1988cz}
M.~S.~Morris and K.~S.~Thorne,
``Wormholes in space-time and their use for interstellar travel: A tool for teaching general relativity,''  Am.\ J.\ Phys.\  {\bf 56} (1988) 395. \doi{10.1119/1.15620}.

\bibitem{Martin-Moruno:2013wfa}
P.~Martin-Moruno and M.~Visser,
``Semiclassical energy conditions for quantum vacuum states,''
JHEP {\bf 1309} (2013) 050, \doi{10.1007/JHEP09(2013)050},
arXiv:\arxiv{1306.2076}[gr-qc].

\bibitem{Lobo:2017oab}
F.~S.~N.~Lobo,
``Wormholes, Warp Drives and Energy Conditions,''
Fundam.\ Theor.\ Phys.\  {\bf 189} (2017). Springer International Publishing, ISBN: 978-3-319-55181-4 \doi{10.1007/978-3-319-55182-1}.

  \bibitem{Kuhfittig:2018voi}
P.~K.~F.~Kuhfittig, ``Traversable wormholes sustained by an extra spatial dimension,''
Phys.\ Rev.\ D {\bf 98} (2018) 064041, \doi{10.1103/PhysRevD.98.064041},
  arXiv: \arxiv{1809.01993}[gr-qc].
  
\bibitem{Hammad:2018ydd}
F.~Hammad, É.~Massé and P.~Labelle,
``Revisiting wormhole energy conditions in Riemann-Cartan spacetimes and under Weyl transformations,''
Phys.\ Rev.\ D {\bf 98} (2018) 124010, \doi{10.1103/PhysRevD.98.124010},
  arXiv:\arxiv{1812.05318}[gr-qc].
  
  \bibitem{Garattini:2007fe}
  R.~Garattini,
 ``Self sustained traversable wormholes and the equation of state,''
  Class.\ Quant.\ Grav.\  {\bf 24} (2007) 1189-1210, \doi{ 	10.1088/0264-9381/24/5/009},
  arXiv:\arxiv{gr-qc/0701019}.
  
  \bibitem{Barros:2018lca}
B.~J.~Barros and F.~S.~N.~Lobo,
``Wormhole geometries supported by three-form fields,''
Phys.\ Rev.\ D {\bf 98} (2018), 044012, \doi{10.1103/PhysRevD.98.044012},
 arXiv:\arxiv{1806.10488}[gr-qc].
 
 \bibitem{Bhar:2016vdn}
  P.~Bhar, F.~Rahaman, T.~Manna and A.~Banerjee,
  ``Wormhole supported by dark energy admitting conformal motion,''
  Eur.\ Phys.\ J.\ C {\bf 76} (2016), 708, \doi{10.1140/epjc/s10052-016-4547-1},
  arXiv:\arxiv{1612.04669}[gr-qc].
  
  \bibitem{Eiroa:2008hv}
E.~F.~Eiroa, M.~G.~Richarte and C.~Simeone,
``Thin-shell wormholes in Brans-Dicke gravity,''
Phys. Lett. A \textbf{373} (2008), 1-4
[erratum: Phys. Lett. \textbf{373} (2009), 2399-2400],
\doi{10.1016/j.physleta.2008.10.065},
arXiv:\arxiv{0809.1623}[gr-qc].


\bibitem{Richarte:2007zz}
M.~G.~Richarte and C.~Simeone,
``Thin-shell wormholes supported by ordinary matter in Einstein-Gauss-Bonnet gravity,''
Phys. Rev. D \textbf{76} (2007), 087502
[erratum: Phys. Rev. D \textbf{77} (2008), 089903],
\doi{10.1103/PhysRevD.77.089903},
arXiv:\arxiv{0710.2041}[gr-qc].

\bibitem{Richarte:2010bd}
M.~G.~Richarte,
``Wormholes and solitonic shells in five-dimensional DGP theory,''
Phys. Rev. D \textbf{82} (2010), 044021,
\doi{10.1103/PhysRevD.82.044021},
arXiv:\arxiv{1003.0741}[gr-qc]

\bibitem{Maeda:2008nz}
H.~Maeda and M.~Nozawa,
``Static and symmetric wormholes respecting energy conditions in Einstein-Gauss-Bonnet gravity,''
Phys. Rev. D \textbf{78} (2008), 024005
\doi{10.1103/PhysRevD.78.024005},
arXiv:\arxiv{0803.1704}[gr-qc].

  \bibitem{Anabalon:2018rzq}
  A.~Anabalón and J.~Oliva,
  ``Four-dimensional Traversable Wormholes and Bouncing Cosmologies in Vacuum,''
  JHEP {\bf 1904} (2019) 106, \doi{10.1007/JHEP04(2019)106},
  arXiv:\arxiv{1811.03497}[hep-th].
  
  \bibitem{Rahaman:2014dpa}
  F.~Rahaman, I.~Karar, S.~Karmakar and S.~Ray,
  ``Wormhole inspired by non-commutative geometry,''
  Phys.\ Lett.\ B {\bf 746} (2015) 73, \doi{10.1016/j.physletb.2015.04.048},
  arXiv:\arxiv{1406.3045} [gr-qc].

\bibitem{Abreu:2012fg}
  E.~M.~C.~Abreu and N.~Sasaki,
  ``Noncommutative Wormholes and the Energy Conditions,''
  arXiv:\arxiv{1207.7130}[hep-th].
  
  \bibitem{Kuhfittig:2013ib}
  P.~K.~F.~Kuhfittig,
  ``Mascroscopic wormholes in noncommutative geometry,''
  Int.\ J.\ Pure Appl.\ Math.\  {\bf 89} (2013) 401, \doi{10.12732/ijpam.v89i3.11}, arXiv:\arxiv{1301.0088}[gr-qc].
  
\bibitem{Garattini:2008xz}
  R.~Garattini and F.~S.~N.~Lobo,
  ``Self-sustained traversable wormholes in noncommutative geometry,''
  Phys.\ Lett.\ B {\bf 671} (2009) 1, 146,
  \doi{10.1016/j.physletb.2008.11.064},
  arXiv:\arxiv{0811.0919}[gr-qc].
  
  \bibitem{Myrzakulov:2015kda}
  R.~Myrzakulov, L.~Sebastiani, S.~Vagnozzi and S.~Zerbini,
  ``Static spherically symmetric solutions in mimetic gravity: rotation curves and wormholes,''
  Class.\ Quant.\ Grav.\  {\bf 33} (2016), 125005, \doi{ 	10.1088/0264-9381/33/12/125005},
  arXiv:\arxiv{1510.02284}[gr-qc].
  
\bibitem{Maldacena:2018gjk}
  J.~Maldacena, A.~Milekhin and F.~Popov,
  ``Traversable wormholes in four dimensions,''
  arXiv: \arxiv{1807.04726}[hep-th].
  

\bibitem{Rahaman:2013xoa}
  F.~Rahaman, P.~K.~F.~Kuhfittig, S.~Ray and N.~Islam,
  ``Possible existence of wormholes in the galactic halo region,''
  Eur.\ Phys.\ J.\ C {\bf 74} (2014) 2750, \doi{10.1140/epjc/s10052-014-2750-5},
  arXiv:\arxiv{1307.1237}[gr-qc].
  
\bibitem{Bueno:2017hyj}
  P.~Bueno, P.~A.~Cano, F.~Goelen, T.~Hertog and B.~Vercnocke,
 ``Echoes of Kerr-like wormholes,''
  Phys.\ Rev.\ D {\bf 97} (2018), 024040, 
  \doi{10.1103/PhysRevD.97.024040},
  arXiv:\arxiv{1711.00391}[gr-qc].
  
\bibitem{Olmo:2015bya}
  G.~J.~Olmo, D.~Rubiera-Garcia and A.~Sanchez-Puente,
  ``Geodesic completeness in a wormhole spacetime with horizons,''
  Phys.\ Rev.\ D {\bf 92} (2015),  044047,
  \doi{10.1103/PhysRevD.92.044047},
  arXiv:\arxiv{1508.03272}[gr-qc]. 

\bibitem{Kagramanova:2013mwv}
V.~Diemer and E.~Smolarek,
``Dynamics of test particles in thin-shell wormhole spacetimes,''
Class. Quant. Grav. \textbf{30} (2013), 175014,
\doi{10.1088/0264-9381/30/17/175014},
arXiv:\arxiv{1302.1705}[gr-qc].

\bibitem{Muller:2008zza}
T.~Muller, ``Exact geometric optics in a Morris-Thorne wormhole spacetime,''   Phys.\ Rev.\ D {\bf 77} (2008) 044043.
\doi{10.1103/PHYSREVD.77.044043}.

\bibitem{Mishra:2017yrh}
A.~Mishra and S.~Chakraborty,``On the trajectories of null and timelike geodesics in different wormhole geometries,''
Eur.\ Phys.\ J.\ C {\bf 78} (2018),  374,
\doi{10.1140/epjc/s10052-018-5854-5},
arXiv:\arxiv{1710.06791}.

\bibitem{Gibbons:2015qja}
G.~W.~Gibbons,
``The Jacobi-metric for timelike geodesics in static spacetimes,''
Class.\ Quant.\ Grav.\  {\bf 33} (2016),  025004, \doi{10.1088/0264-9381/33/2/025004},
arXiv:\arxiv{1508.06755}[gr-qc].

\bibitem{Chanda:2016aph}
S.~Chanda, G.~W.~Gibbons and P.~Guha,
``Jacobi-Maupertuis-Eisenhart metric and geodesic flows,''
J.\ Math.\ Phys.\  {\bf 58} (2017), 032503,
\doi{10.1063/1.4978333},
arXiv:\arxiv{1612.00375}[math-ph].

\bibitem{Tsiganov:2001}
A. V.~Tsiganov,
"The Maupertuis Principle and Canonical Transformations of the Extended Phase Space",
J.\ Nonlinear. \ Math. \ Phys, {\bf 8} (2001) no.1, 157-182, 
\doi{10.2991/jnmp.2001.8.1.12}, 
arXiv:\arxiv{nlin/0101061}[nlin.SI].

\bibitem{Chanda:2016sjg}
  S.~Chanda, G.~W.~Gibbons and P.~Guha,
  ``Jacobi–Maupertuis metric and Kepler equation,''
  Int.\ J.\ Geom.\ Meth.\ Mod.\ Phys.\  {\bf 14} (2017) no.07,  1730002,
 \doi{10.1142/S0219887817300021},
arXiv:\arxiv{1612.07395}[math-ph].

  


\bibitem{Das:2016opi}
P.~Das, R.~Sk and S.~Ghosh,``Motion of charged particle in Reissner–Nordström spacetime: a Jacobi-metric approach,''
Eur.\ Phys.\ J.\ C {\bf 77} (2017),  735, 
\doi{10.1140/epjc/s10052-017-5295-6},
arXiv:\arxiv{1609.04577}[gr-qc].

\bibitem{Izquierdo:2002jt}
  A.~Alonso Izquierdo, M.~A.~Gonzalez Leon, J.~Mateos Guilarte and M.~de la Torre Mayado,
``Jacobi metric and Morse theory of dynamical systems,'' Publ. R. Soc. Mat. Esp. {\bf{6}} (2004) 81--91,  arXiv:\arxiv{math-ph/0212017}.
  
  \bibitem{Szydlowski:1996}
  M. ~Szydlowski, Geometry of spaces with the Jacobi metric, Journal of Mathematical Physics 37, 346 (1996),
\doi{10.1063/1.531394}.



  \bibitem{Hochberg:1997wp}
  D.~Hochberg and M.~Visser,
  ``Geometric structure of the generic static traversable wormhole throat,''
  Phys.\ Rev.\ D {\bf 56} (1997) 4745,
  \doi{10.1103/PhysRevD.56.4745},  
  arXiv:\arxiv{gr-qc/9704082}.
  
\bibitem{Pin:1975}  
O.~Chong Pin, "Curvature and mechanics", Advances in Mathematics,{\bf 15},(1975), 3, 269,
\doi{10.1016/0001-8708(75)90139-5}.

\bibitem{Arganaraz:2020}
  M.~Arganaraz and O.~Lasso Andino,
  `` A covariant flare-out condition from a Jacobi metric,'' Work in progress.
  
  \bibitem{Cariglia:2015bla}
  M.~Cariglia and F.~K.~Alves,
  ``The Eisenhart lift: a didactical introduction of modern geometrical concepts from Hamiltonian dynamics,''
  Eur.\ J.\ Phys.\  {\bf 36} (2015),  025018,
 \doi{10.1088/0143-0807/36/2/025018},
  arXiv:\arxiv{1503.07802}.
  
  \bibitem{Finn:2018cfs}
  K.~Finn, S.~Karamitsos and A.~Pilaftsis,
  ``Eisenhart lift for field theories,''
  Phys.\ Rev.\ D {\bf 98} (2018) no.1,  016015,
  \doi{10.1103/PhysRevD.98.016015},
  arXiv:\arxiv{1806.02431}[physics.class-ph].
  

\bibitem{Harko:2013yb}
  T.~Harko, F.~S.~N.~Lobo, M.~K.~Mak and S.~V.~Sushkov,
  ``Modified-gravity wormholes without exotic matter,''
  Phys.\ Rev.\ D {\bf 87} (2013) no.6,  067504
  doi: \doi{10.1103/PhysRevD.87.067504}
  arXiv:\arxiv{1301.6878}[gr-qc].
  
  \bibitem{Chanda:2019guf}
  S.~Chanda, G.~W.~Gibbons, P.~Guha, P.~Maraner and M.~C.~Werner,
  ``Jacobi-Maupertuis Randers-Finsler metric for curved spaces and the gravitational magnetoelectric effect,''
  arXiv:\arxiv{1903.11805} [gr-qc].
  
  \bibitem{Boonserm:2018orb}
  P.~Boonserm, T.~Ngampitipan, A.~Simpson and M.~Visser,
  ``Exponential metric represents a traversable wormhole,''
  Phys.\ Rev.\ D {\bf 98} (2018),  084048,
  \doi{10.1063/1.5098869},
  arXiv: \arxiv{1805.03781}[gr-qc].
  
  \bibitem{Damour:2007ap}
  T.~Damour and S.~N.~Solodukhin,
  ``Wormholes as black hole foils,''
  Phys.\ Rev.\ D {\bf 76} (2007) 024016,
  \doi{10.1103/PhysRevD.76.024016},
  arXiv: \arxiv{0704.2667}[gr-qc].
  
  \bibitem{Kim:2001ri}
  S.~W.~Kim and H.~Lee,
  ``Exact solutions of a charged wormhole,''
  Phys.\ Rev.\ D {\bf 63} (2001) 064014,
  \doi{10.1103/PhysRevD.63.064014},
  arXiv:\arxiv{gr-qc/0102077}.
  
\bibitem{Kuhfittig:2011xh}
  P.~K.~F.~Kuhfittig,
  ``On the feasibility of charged wormholes,''
  Central Eur.\ J.\ Phys.\  {\bf 9} (2011) 1144,
  \doi{10.2478/s11534-011-0043-2},
  arXiv:\arxiv{1104.4662}[gr-qc].


\end{thebibliography}
\end{document}